\documentclass[%
superscriptaddress,
showpacs,
nofootinbib,
amsmath,amssymb,
prc,
twocolumn,
floatfix ]%
{revtex4-1}

\usepackage{color}
\usepackage{CJK}

\usepackage{graphicx}
\usepackage{dcolumn}
\usepackage{bm}
\usepackage{hyperref}

\allowdisplaybreaks

\begin{document}

\begin{CJK*}{UTF8}{}
\title{Exotic nonaxial-octupole shapes in $N=184$ isotones from covariant density functional theories}
\CJKfamily{gbsn}
\author{Jie Zhao (赵杰)}%
\affiliation{Center for Circuits and Systems, Peng Cheng Laboratory, Shenzhen 518055, China}
\author{Zheng-Gang Wu (巫政钢)}%
\affiliation{Center for Circuits and Systems, Peng Cheng Laboratory, Shenzhen 518055, China}

\date{\today}

\begin{abstract}
The nonaxial octupole shape in some nuclei with $N = 184$, namely, 
$^{284}$Fm, $^{286}$No, $^{288}$Rf, and $^{290}$Sg, is investigated using covariant density functional theories.
Employing the density-dependent point-coupling covariant density functional theory with the parameter set DD-PC1
in the particle-hole channel, it is found that the ground states of $^{284}$Fm, $^{286}$No, $^{288}$Rf, and $^{290}$Sg
have pure nonaxial octupole shapes with deformation parameters $\beta_{31} \approx 0.08$ and $\beta_{33} \approx -0.01 \sim -0.03$.
The energy gain due to the $\beta_{31}$ and $\beta_{33}$ distortion is $\sim$ 1 MeV. 
The occurrence of the nonaxial octupole correlations is mainly from the proton orbitals $1i_{13/2}$ and $2f_{7/2}$, 
which are close to the proton Fermi surface.
The dependence of the nonaxial octupole effects on the form of the energy density functional and on the parameter set is also studied.
\end{abstract}

\maketitle

\end{CJK*}

\bigskip

\section{Introduction~\label{sec:Introduction}}
The occurrence of spontaneous symmetry breaking leads to the intrinsic shape 
of many atomic nuclei deviating from a sphere.
The deformation of a nucleus is usually described by a multipole expansion of the 
nuclear surface 
\begin{equation}
	R ( \theta, \varphi ) = 
	R_0 \left[  1 + 
	\beta_{00} + 
	\sum_{\lambda=1}^{\infty} \sum_{\mu=-\lambda}^\lambda 
	\beta_{\lambda \mu}^* Y_{\lambda \mu} ( \theta, \varphi )    
	\right] ,
	\label{Eq:SurfaceDeformation}
\end{equation}
where $\beta_{\lambda\mu}$'s are deformation parameters.
The presence of quadrupole deformations $\beta_{20}$, $\beta_{22}$ is well known 
and the related nuclear phenomena have been studied extensively.
Recently, enhanced reduced electric-octupole transition probabilities, $B(E3)$, 
were observed in $^{224}$Ra~\cite{Gaffney2013_Nature497-199}, $^{144}$Ba~\cite{Bucher2016_PRL116-112503},
$^{146}$Ba~\cite{Bucher2017_PRL118-152504}, and $^{228}$Th~\cite{Chishti2020_NP16-853},
providing direct experimental evidence of static octupole deformation. 
In addition, the octuple correlations between multiple chiral doublet 
bands in $^{78}$Br~\cite{Liu2016_PRL116-112501} and the coexistence of chirality 
and octupole correlations in $^{76}$Br~\cite{Xu2022_PLB833-137287} have also been observed.

Besides the axial octupole deformation $\beta_{30}$, the nonaxial octupole deformations $\beta_{31}$,
$\beta_{32}$, and $\beta_{33}$ are also of particular interests.
Among these three nonaxial octupole deformations, the $\beta_{32}$ deformation have attracted much more attention 
due to its special symmetry properties, i.e., a nucleus with a pure $\beta_{32}$ deformation 
($\beta_{\lambda\mu}=0$ if $\lambda\ne3$ and $\mu\ne2$) has a tetrahedral shape with the symmetry group $T_d^D$.
The predicted shell gaps at specific proton or neutron numbers for a nucleus with tetrahedral symmetry are comparable 
or even stronger than those at spherical shapes~\cite{Li1994_PRC49-R1250,Dudek2002_PRL88-252502,Dudek2007_IJMPE16-516,
	Dudek2003_APPB34-2491,Heiss1999_PRC60-034303,Arita2014_PRC89-054308}.
Thus there may be a static tetrahedral shape or strong tetrahedral 
correlations for a nucleus with such proton or neutron numbers.

Various nuclei were predicted to have ground or isomeric states with tetrahedral shapes  
from the macroscopic-microscopic (MM) model 
\cite{Dudek2002_PRL88-252502,Dudek2007_IJMPE16-516,Dudek2006_PRL97-072501,%
	Schunck2004_PRC69-061305R,Dudek2014_PS89-054007,Jachimowicz2017_PRC95-034329}
and the Skyrme Hartree-Fock (SHF) model, 
the SHF plus BCS model, or the Skyrme Hartree-Fock-Bogoliubov (SHFB) model
\cite{Dudek2007_IJMPE16-516,Schunck2004_PRC69-061305R,Dudek2006_PRL97-072501,%
	Yamagami2001_NPA693-579,Olbratowski2006_IJMPE15-333,Zberecki2009_PRC79-014319,%
	Zberecki2006_PRC74-051302R,Takami1998_PLB431-242}.
The ground state shapes of even-even Zr isotopes were also studied within the 
multidimensionally constrained relativistic Hartree-Bogoliubov (MDC-RHB) 
model~\cite{Zhao2017_PRC95-014320}, 
possible tetrahedral shapes in the ground and isomeric states were predicted 
for nucleus around $^{110}$Zr. 
Consistent results were obtained from three-dimensional (3D) lattice calculations~\cite{Xu2023_arXiv}.
The rotational properties of tetrahedral nuclei have also been studied theoretically 
\cite{Gao2004_CPL21-806,Tagami2013_PRC87-054306,Tagami2015_JPG42-015106,
	Chen2010_NPA834-378c,Chen2013_NPR30-278,Bijker2014_PRL112-152501,Tagami2018_PRC98-024304}.
Several experiments were devoted to the study of tetrahedral shapes in 
$^{160}$Yb \cite{Bark2010_PRL104-022501},
$^{154,156}$Gd \cite{Bark2010_PRL104-022501,Jentschel2010_PRL104-222502,Doan2010_PRC82-067306},
$^{230,232}$U \cite{Ntshangase2010_PRC82-041305R}, 
and $^{108}$Zr \cite{Sumikama2011_PRL106-202501}.

More interestingly, although the ground state of the well known magic nuclei $^{16}$O 
is predicted to have spherical shape from mean field calculations, 
the restoration of the rotational and parity symmetry leads to the 
occurrence of the tetrahedral symmetry~\cite{Wang2019_PLB790-498}.
This is consistent with the very recent predictions from the algebraic cluster model~\cite{Bijker2014_PRL112-152501}
and the $ab$ $initio$ lattice calculation in the framework of nuclear lattice effective field theory~\cite{Epelbaum2014_PRL112-102501}.
Similar projection-after-variation calculations for $^{96}$Zr were also presented in Ref.~\cite{Rong2023_PLB840-137896}.
The $\beta_{32}$ deformation also shows up after symmetry restorations, although this deformation parameter 
was predicted to be zero from the mean-field calculations.
This may indicate that the non-axial octupole $\beta_{32}$ deformation plays much more important 
role in the ground and low-lying states of nuclear many-body system than expected.

Compared to $\beta_{32}$, less attention has been paid to the distortion effect from
$\beta_{31}$ and $\beta_{33}$, although still sizable shell gaps 
can be found in the single-particle levels at specific proton or neutron numbers when 
$\beta_{31}$ or $\beta_{33}$ are considered~\cite{Hamamoto1991_ZPD21-163,Dudek2002_PRL88-252502}.
From the MM model, some nuclei in Se, Ba, Ra isotopic chains were predicted to have ground 
or isometric states with nonzero $\beta_{31}$ or $\beta_{33}$ together with sizable quadrupole 
deformations~\cite{Liu2018_CPC42-074105}.
Recently, the nuclear octupole fourfold neutron magic number $N=136$ 
and 196 were introduced. From the potential energy surfaces generated from MM model, 
several nuclei around this range are predicted to have equilibrium shapes with
nonzero $\beta_{31}$ or $\beta_{33}$~\cite{Yang2022_PRC105-034348,Yang2022_PRC106-054314,
	Yang2023_PRC107-054304,Jachimowicz2017_PRC95-034329}.
The ground state shapes of $^{64}$Ge and $^{68}$Se are predicted to have nonzero $\beta_{33}$
from SHF plus BCS calculations~\cite{Takami1998_PLB431-242}.
A fully 3D symmetry-unrestricted SHFB calculations indicate that  
the oblate ground state of $^{68}$Se is unstable against the triangular $\beta_{33}$ distortion,
the predicted ground state have equilibrium shape with 
$\beta_{20}=-0.28$ and $\beta_{33}=0.08$~\cite{Yamagami2001_NPA693-579}. 

In recent years, nuclei with $Z \simeq 100$ have been studied extensively 
because such studies can not only reveal the structure for these nuclei 
but also give useful structure information for superheavy nuclei.
For example the observed very low-lying $2^{-}$ bands in several 
$N=150$ isotones were attributed to the $Y_{32}$ correlations~\cite{Chen2008_PRC77-061305R}.
The nuclear shapes with a non-zero $\beta_{32}$ superposed on a sizable $\beta_{20}$
were also predicted within MDC-RHB model for these $N=150$ isotones~\cite{Zhao2012_PRC86-057304}.

However, the $V_{4}$ symmetry is imposed in the MDC-RHB model, only the shape degrees of 
freedom $\beta_{\lambda\mu}$ with even $\mu$ are allowed~\cite{Lu2012_PRC85-011301R,Lu2014_PRC89-014323,
	Zhou2016_PS91-063008,Zhao2017_PRC95-014320}.
To include the $\beta_{31}$ and $\beta_{33}$ deformations, 
we released the constraint of the $V_{4}$ symmetry and extended the MDC-RHB model 
by imposing only the simplex-$y$ symmetry.
The RHB equation is solved by expanding the Dirac spinors in simplex-$y$ harmonic oscillator (HO) basis.
Thus all four magnetic components of the octpole deformation 
$\beta_{3\mu}$, $\mu=0,1,2,3$ can be present simultaneously.   
In this work, we present the microscopic investigation of the $\beta_{31}$ and $\beta_{33}$
distortion effect on $N=184$ isotones within the extended MDC-RHB model.
The theoretical framework and method are introduced in Sec.~\ref{sec:model}.
The results for $N=184$ isotones are described and discussed in Sec~\ref{sec:results}. 
Sec.~\ref{sec:summary} contains a summary of the principal results.

\section{\label{sec:model}Theoretical framework}
In covariant density functional theory, one can derive the Dirac-Hartree-Bogoliubov equation within the Green's
function technique~\cite{Kucharek1991_ZPA339-23,Ring1996_PPNP37-193}, 
the obtained RHB equation reads
\begin{equation}
	\label{eq:rhb}
	\int d^{3}\bm{r}^{\prime}
	\left(\begin{array}{cc} h_{D}-\lambda & \Delta \\ -\Delta^{*} & -h_{D}+\lambda \end{array}\right)
	\left(\begin{array}{c} U_{k} \\ V_{k} \end{array}\right)
	= E_{k}\left(\begin{array}{c} U_{k} \\ V_{k} \end{array}\right),
\end{equation}
where $E_{k}$ is the quasiparticle energy, 
$\left( U_k, \ V_{k} \right)^\mathrm{T}$ is the quasiparticle wave function,
$\lambda$ is the chemical potential, 
and $\hat{h}_{D}$ is the single-particle Dirac Hamiltonian
\begin{equation}
	\label{eq:hamiltonian}
	\hat{h}_{D} = \bm{\alpha}\cdot[\bm{p}-\bm{V}(\bm{r})] 
	+ \beta[M+S(\bm{r})]
	+ V_{0}(\bm{r}) + \Sigma_R(\bm{r}),
\end{equation}
$S(\bm{r})$, $\bm{V}(\bm{r})$, and $\Sigma_R(\bm{r})$ denote 
the scalar potential, vector potential, and rearrangement terms respectively.
The pairing potential is given by
\begin{equation}
	\begin{aligned}
		\Delta_{p_{1}p_{2}}(\bm{r}_{1}\sigma_{1},\bm{r}_{2}\sigma_{2})
		= & \int d^{3}\bm{r}_{1}^{\prime} d^{3}\bm{r}_{2}^{\prime} 
		\sum_{\sigma_{1}^{\prime}\sigma_{2}^{\prime}}^{p_{1}^{\prime}p_{2}^{\prime}} \\
		& V_{p_{1}p_{2},p_{1}^{\prime}p_{2}^{\prime}}^{{\rm pp}} 
		(\bm{r}_{1}\sigma_{1}, \bm{r}_{2}\sigma_{2},
		\bm{r}_{1}^{\prime}\sigma_{1}^{\prime}, \bm{r}_{2}^{\prime}\sigma_{2}^{\prime}) \\
		& \times \kappa_{p_{1}^{\prime}p_{2}^{\prime}} 
		(\bm{r}_{1}^{\prime}\sigma_{1}^{\prime}, 
		\bm{r}_{2}^{\prime}\sigma_{2}^{\prime}),
	\end{aligned}
\end{equation}
where $p=f,g$ is used to represent the large and small components of the Dirac spinors. 
$V^{{\rm pp}}$ is the effective pairing interaction
and $\kappa(\bm{r}_{1}\sigma_{1},\bm{r}_{2}\sigma_{2})$ is the pairing tensor.
As is usually done in the RHB theory, only the large components of the spinors are 
used to build the pairing potential~\cite{Serra2002_PRC65-064324}. 
In the pp channel, we use a separable pairing force of finite range \cite{Tian2009_PLB676-44, 
	Tian2009_PRC80-024313, Niksic2010_PRC81-054318}
\begin{equation}
	\label{eq:separable}
	V( \bm{r}_1 - \bm{r}_2 ) = -{G} \delta( \tilde{\bm{R}} - \tilde{\bm{R}^\prime} )
	P(\tilde{\bm{r}}) P(\tilde{\bm{r}^\prime}) \frac{1-\hat{P}_\sigma}{2} ,
\end{equation}
where $G$ is the pairing strength and $\tilde{\bm{R}}$ and $\tilde{\bm{r}}$ 
are the center-of-mass and relative coordinates of the two nucleons, respectively.
$P(\bm{r})$ denotes the Gaussian function,
\begin{equation}
	P(\bm{r}) = \frac{1}{(4\pi a^2)^{3/2}} e^{-{r^2}/{4 a^2}},
\end{equation}
where $a$ is the effective range of the pairing force.
The two parameters $G=728$ MeV$\cdot$fm$^3$ and $a=0.644$ fm 
\cite{Tian2009_PLB676-44, Tian2009_PRC80-024313}
have been adjusted to reproduce the density dependence of the 
pairing gap at the Fermi surface
in symmetric nuclear matter and calculated with the Gogny force D1S.

The RHB equation~(\ref{eq:rhb}) is solved by expanding the large and small components
of the spinors $U_{k}(\bm{r}\sigma)$ and $V_{k}(\bm{r}\sigma)$ in 
simplex-$y$ HO basis, where the 
axially deformed (AD) HO basis states are used to build 
the eigenfunctions of the simplex-y operator $S_{y} = Pe^{-i\pi j_{y}}$.
$P$ denotes the parity operator.
The eigenstates of the $S_{y}$ operator with eigenvalues $s=+i$ and $s=-i$ 
reads~\cite{Bjelcic2020_CPC253-107184}
\begin{equation}
	\left|n_{z} n_{r} \Lambda; s=+i \right\rangle = \frac{1}{\sqrt{2}} 
	\left( i\left|n_{z} n_{r} \Lambda; 1/2 \right\rangle 
	+ \left|n_{z} n_{r} -\Lambda; -1/2 \right\rangle \right),
\end{equation}
\begin{equation}
	\left|n_{z} n_{r} \Lambda; s=-i \right\rangle = \frac{1}{\sqrt{2}} 
	\left( \left|n_{z} n_{r} \Lambda; 1/2 \right\rangle 
	+ i \left|n_{z} n_{r} -\Lambda; -1/2 \right\rangle \right),
\end{equation}
where $\left|n_{z} n_{r} \Lambda; m_{s} \right\rangle$ refers to the eigenfunctions of the ADHO potential
and $n_{z}$, $n_{r}$, $\Lambda$, and $m_{s}$ are the corresponding quantum numbers.
These states are related by the time-reversal operator 
\begin{equation}
	\mathcal{T} \left|n_{z} n_{r} \Lambda; s=\pm i \right\rangle = \mp \left|n_{z} n_{r} \Lambda; s=\mp i \right\rangle.
\end{equation}
Due to the time-reversal symmetry, the RHB matrix is block-diagonalized into 
two smaller ones denoted by the quantum number $S_{y}=\pm i$, respectively. 
For a system with the time-reversal symmetry, 
it is only necessary to diagonalize the matrix with $S_{y}=+i$, and 
the other half is obtained by applying the time reversal operation to Dirac spinors.
We should note that the large and small components of the Dirac spinors are expanded 
in the simplex-$y$ eigenfunctions of opposite eigenvalues.
Within the simplex-$y$ symmetry, the deformations corresponding to the four magnetic components of 
the octupole deformation $\beta_{3\mu}$, $\mu=0,1,2$, and 3 can be present 
simultaneously~\cite{Dobaczewski1997_CPC102-166}.

In practical calculations, the ADHO basis is truncated as
$\left[ n_{z} / Q_{z} + (2n_{\rho}+|m|) / Q_{\rho} \right] \leq N_{f}$
\cite{Warda2002_PRC66-014310,Lu2014_PRC89-014323},
for the large component of the Dirac spinor. 
$N_{f}$ is a certain integer constant and $Q_{z}=\max(1,b_{z}/b_{0})$
and $Q_{\rho}=\max(1,b_{\rho}/b_{0})$ are constants calculated from
the oscillator lengths $b_{0}=1/\sqrt{M\omega_0}$, $b_{z}=1/\sqrt{M\omega_z}$, 
and $b_{\rho}=1/\sqrt{M\omega_\rho}$.
$\omega_z$ and $\omega_{\rho}$ are defined through relations 
$\omega_{z}=\omega_{0}\exp(-\sqrt{5/4\pi}\beta_{{\rm B}})$
and $\omega_{\rho}=\omega_{0}\exp(\sqrt{5/16\pi}\beta_{{\rm B}})$,
where $\omega_{0}=(\omega_{z}\omega_{\rho}^{2})^{1/3}$ is
the frequency of the corresponding spherical oscillator and 
$\beta_{{\rm B}}$ is the deformation of the basis. 
For the small component, the truncation is made up to $N_g=N_f+1$ major 
shells in order to avoid spurious states \cite{Gambhir1990_APNY198-132}.
$N_{f}=20$ is adopted in the present calculations.
The deformation parameter $\beta_{\lambda\mu}$ is obtained from the corresponding
multipole moment using
\begin{equation}
	\beta_{\lambda\mu}^{\tau}=\frac{4\pi}{3N_{\tau}R^{\lambda}}Q_{\lambda\mu}^{\tau},
\end{equation}
where $R=1.2 \times A^{1/3}$ fm
and $N_{\tau}$ is the number of proton, neutron, or nucleons.
For the details of the RHB model and the ADHO basis, 
we refer the readers to Ref~\cite{Zhao2017_PRC95-014320}.

\section{\label{sec:results}Results and discussions}
\begin{figure}[h]
	\begin{center}
		\includegraphics[width=0.45\textwidth]{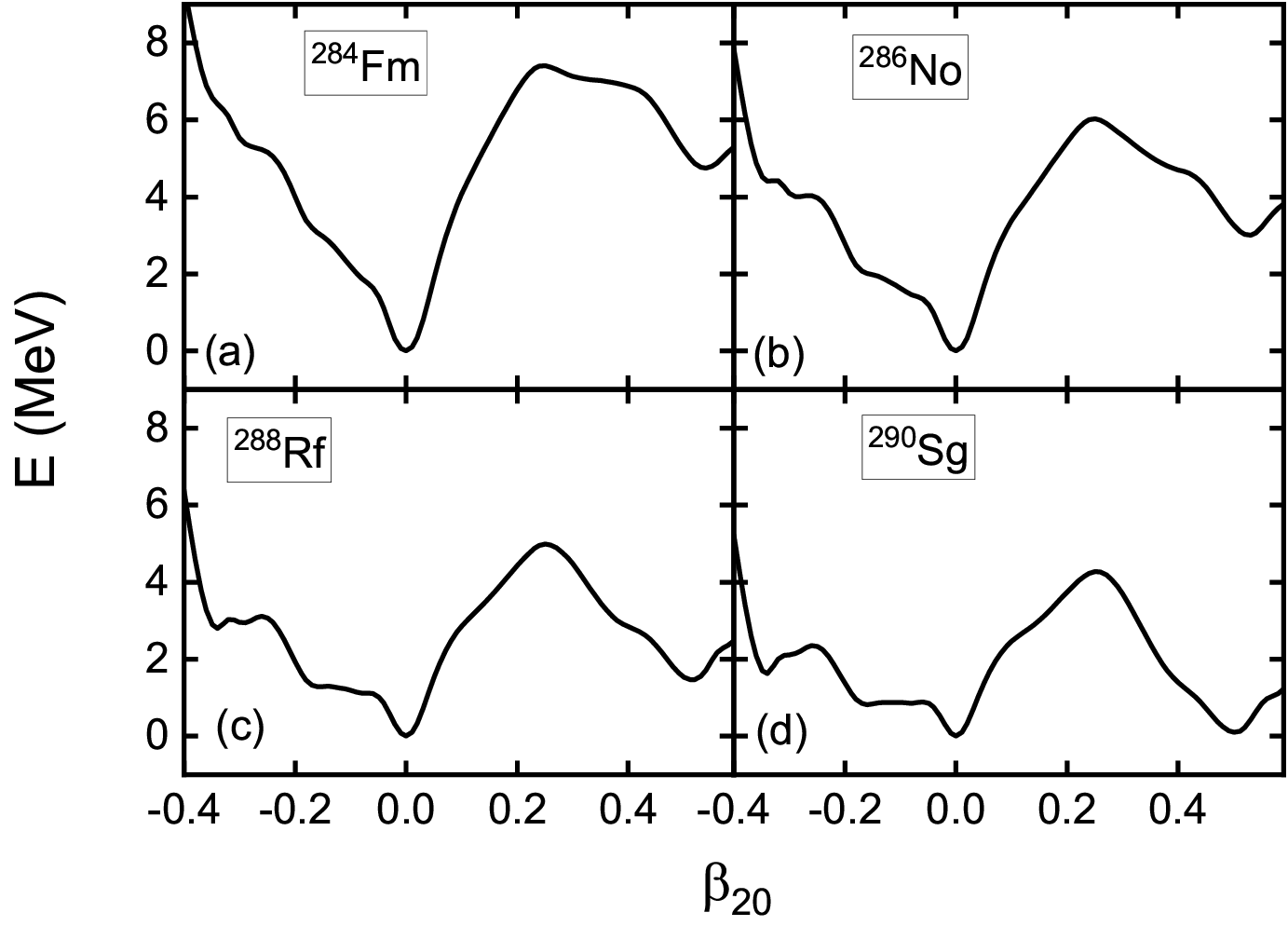}
	\end{center}
	\caption{(Color online) 
		Potential energy curves ($E \sim \beta_{20}$) for $N=184$ isotones 
		$^{284}$Fm, $^{286}$No, $^{288}$Rf, and $^{290}$Sg.
		The energy is normalized with respect to the energy minimum with $\beta_{20} = 0$.
		Axial symmetry and reflection symmetry is imposed.
		The functional DD-PC1 is used in the RHB calculations.}
	\label{fig:TTm}
\end{figure}

In Fig.~\ref{fig:TTm}, we display the calculated one-dimensional (1D) 
potential energy curves ($E \sim \beta_{20}$) for even-even $N=184$ 
nuclei $^{284}$Fm, $^{286}$No, $^{288}$Rf, and $^{290}$Sg when axial 
and reflection symmetry imposed. 
Calculations were performed with parameter set DD-PC1~\cite{Niksic2008_PRC78-034318}. 
The ground state shape of these four
nuclei investigated here are all predicted to be spherical when nuclear 
shapes are restricted to be axial and reflection symmetric.
Additionally, we observe a prolate minimum with $\beta_{20} \approx 0.5$.
In the case of $^{284}$Fm, the energy of the prolate minimum is approximately 4.7 MeV higher 
than the spherical minimum. As the proton number increases, the energy of the
prolate minimum decreases. For $^{290}$Sg, the energy of this minimum is 
only about 0.09 MeV higher than the corresponding spherical minimum.

\begin{figure}[htb]
	\begin{center}
		\includegraphics[width=0.45\textwidth]{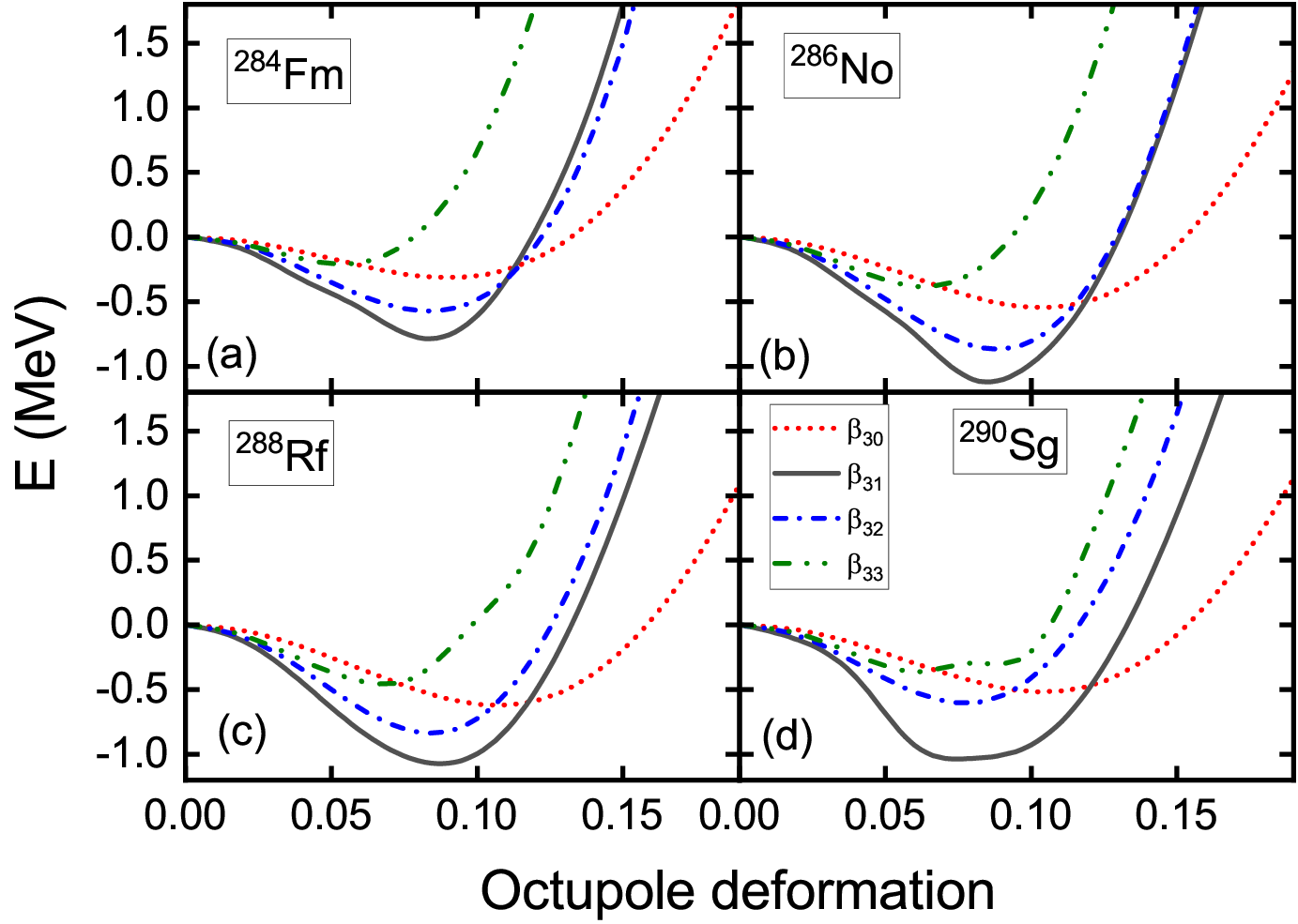}
	\end{center}
	\caption{(Color online) 
		The binding energy $E$ (relative to the energy of the spherical shape
		$\beta_{20}=0$ and $\beta_{3\mu} =0$, $\mu=0,1,2,3$) for $N=184$ isotones $^{284}$Fm, 
		$^{286}$No, $^{288}$Rf, and $^{290}$Sg as a function of the octupole 
		deformation parameter $\beta_{30}$ (dotted line), $\beta_{31}$ (solid line),
		$\beta_{32}$ (dash-dotted line), and $\beta_{33}$ (dash-dot-dotted line).
		The functional DD-PC1 is used in the RHB calculations.
	}
	\label{fig:Octu}
\end{figure}

To explore the impact of axial octupole deformation $\beta_{30}$ and nonaxial 
octupole deformation $\beta_{3\mu}$, $\mu=1,2,3$, on the ground state properties 
along the N = 184 isotonic chain, we perform 1D constrained calculations 
around the minimum $\beta_{20} \approx 0$ and obtain potential energy curves, 
i.e., the total binding energy as a function of $\beta_{3\mu}$, $\mu=0,1,2,3$.
At each point on a potential energy curve, the energy is automatically minimized
with respect to other shape degrees of freedom, such as $\beta_{20}$, $\beta_{22}$, 
$\beta_{3\nu}$ for $\nu \neq \mu$, $\beta_{40}$ etc.
In Fig.~\ref{fig:Octu}, we show obtained potential energy curves for these $N=184$ isotones.
The total binding energy (relative to the energy at $\beta_{20}=0$, $\beta_{3\mu}=0$) as a 
function of $\beta_{30}$ is shown with dotted line, while the one as a function of $\beta_{31}$,
$\beta_{32}$, and $\beta_{33}$ is shown with solid line, dash-dotted line, and dash-dot-dotted 
line respectively.
The effect of axial octupole deformation $\beta_{30}$ on the ground states of these $N=184$ 
nuclei has already been discussed in Ref.~\cite{Agbemava2016_PRC93-044304}.
As illustrated in Fig.~\ref{fig:Octu} (dotted line), the inclusion of $\beta_{30}$ lowers 
the energy, and a minimum around $\beta_{30} = 0.1$ develops for all four nuclei.
The energy gain due to the inclusion of $\beta_{30}$ is about 0.3 MeV for $^{284}$Fm.
As the proton number increases, the energy gain due to $\beta_{30}$ distortion rises,
reaching a maximum for $^{288}$Rf, where the value is about 0.6 MeV.
This energy gain decreases to 0.5 MeV for $^{290}$Sg.

\begin{table}
	\centering
	\caption{\label{tab:oct} %
		The axial octupole deformation $\beta_{30}$ and nonaxial octupole deformation parameters
		$\beta_{31}$, $\beta_{32}$, and $\beta_{33}$ together with binding energies 
		(relative to the energy of the spherical shape $\beta_{20}=0$, $\beta_{3\mu}=0$, $\mu=0,1,2,3$) 
		for various energy minima with $\beta_{3\mu} \neq 0$ in $N=184$ nuclei calculated with parameter
		set DD-PC1. All energies are in MeV.
	}
	\footnotesize
	\begin{tabular}{lcccccc}
		\hline\hline
		Nucleus	   & $\beta_{30}$ & $\beta_{31}$ & $\beta_{32}$ & $\beta_{33}$ & $E_{\rm{depth}}$ \\ \hline
		$^{284}$Fm & 0.086        & 0.000        & 0.000        & 0.000        & 0.310            \\ 
		& 0.000        & 0.077        & 0.000        & 0.042        & 0.779            \\ 
		& 0.000        & 0.000        & 0.080        & 0.000        & 0.569            \\
		& 0.000        & 0.006        & 0.000        & 0.051        & 0.207            \\ 
		$^{286}$No & 0.100        & 0.000        & 0.000        & 0.000        & 0.543            \\ 
		& 0.000        & 0.078        & 0.000        & $-0.016$     & 0.907            \\ 
		& 0.000        & 0.000        & 0.086        & 0.000        & 0.865            \\
		& 0.000        & 0.006        & 0.000        & 0.061        & 0.381            \\ 
		$^{288}$Rf & 0.104        & 0.000        & 0.000        & 0.000        & 0.619            \\ 
		& 0.000        & 0.084        & 0.000        & $-0.008$     & 1.062            \\ 
		& 0.000        & 0.000        & 0.083        & 0.000        & 0.835            \\
		& 0.000        & 0.007        & 0.000        & 0.065        & 0.449            \\
		$^{290}$Sg & 0.101        & 0.000        & 0.000        & 0.000        & 0.515            \\ 
		& 0.000        & 0.081        & 0.000        & $-0.031$     & 1.033            \\ 
		& 0.000        & 0.000        & 0.075        & 0.000        & 0.602            \\
		& 0.000        & 0.009        & 0.000        & 0.064        & 0.371            \\
		\hline\hline
	\end{tabular}
\end{table}

The primary focus of this manuscript is on the nonaxial octupole deformations
$\beta_{31}$, $\beta_{32}$, and $\beta_{33}$.
Among these three nonaxial octupole deformations, the effect of $\beta_{32}$ distortion has been 
extensively studied due to its symmetry properties. The pure $\beta_{32}$ deformation 
is known as the tetrahedral shape, where the single-particle levels split into multiplets 
with degeneracies equal to the irreducible representations of the $T_{d}^{D}$ group.
Consequently, large energy gaps can be obtained at certain proton or neutron numbers such as 
$Z/N = 20, 40, 70$ etc.~\cite{Zhao2017_PRC95-014320}.
As depicted in Fig.~\ref{fig:Octu} (dash-dotted line), when $\beta_{32}$ is constrained,
the energy of these four $N=184$ nuclei decreases as $\beta_{32}$ increases, 
reaching a minimum at $\beta_{32} \approx 0.08$.
The energy gain due to $\beta_{32}$ distortion is about 0.57 MeV for $^{284}$Fm,
and values are approximately 0.87, 0.84, and 0.60 MeV for $^{286}$No, $^{288}$Rf, and $^{290}$Sg respectively.
Clearly, the effect of $\beta_{32}$ distortion is lager than that of $\beta_{30}$ for these nuclei.

The impact of nonaxial octupole deformations $\beta_{31}$ and $\beta_{33}$ on the ground state 
properties has been less studied microscopically in the previous works.
In Fig.~\ref{fig:Octu} (solid line), potential energy curves obtained by constraining $\beta_{31}$ are shown.
Interestingly, for all four nuclei studied here, the deepest minima were obtained when $\beta_{31}$ was constrained.
It is important to noted that when $\beta_{31}$ is constrained, only the variationally determined $\beta_{33}$ value is not zero.
Taking $^{284}$Fm as an example, this minimum is located at $\beta_{31} = 0.077$ and $\beta_{33} = 0.042$, 
resulting in an energy gain of 0.779 MeV. 
For $^{286}$No, the minimum is located at $\beta_{31} = 0.078$ and $\beta_{33} = -0.016$, 
with an energy gain of 0.907 MeV. 
In the case of $^{288}$Rf, the depth of this minimum is 1.062 MeV 
characterized by $\beta_{31} = 0.084$ and $\beta_{33} = -0.008$.
Finally, for $^{290}$Sg, the energy gain due to $\beta_{31}$ and $\beta_{33}$ distortion is 1.033 MeV,
occurring at $\beta_{31} = 0.081$ and $\beta_{33} = -0.031$.
We note that the energy curve as a function of the $\beta_{31}$ deformation 
[solid line in Fig.~\ref{fig:Octu} (d)] is rather soft around the minimum for $^{290}$Sg,
this softness suggests that certain dynamical correlations beyond the mean-field approximation are important.

Similarly, when we constrain $\beta_{33}$ and variationally determine the other deformation parameters 
with the initial values set to zero, we obtain the 1D potential energy curve shown in 
Fig.~\ref{fig:Octu} dash-dot-dotted line.
As $\beta_{33}$ increases, the energy first decreases and then increases, 
and leads to a shallow minimum at finite $\beta_{33}$.
The depth of this minimum is the smallest when 
compared to the cases where we constrained $\beta_{30}$, $\beta_{32}$ or $\beta_{31}$.
Thus remarkable $\beta_{31}$ deformation together with moderate $\beta_{33}$ deformation  
were predicted for the ground states of $^{284}$Fm, $^{286}$No, $^{288}$Rf, and $^{290}$Sg.
We summarized the values of the octupole deformation parameters and the energy gain 
due to $\beta_{3\mu}$ distortion for various minima discussed above in Table~\ref{tab:oct}.

\begin{figure}[h]
	\begin{center}
		\includegraphics[width=0.45\textwidth]{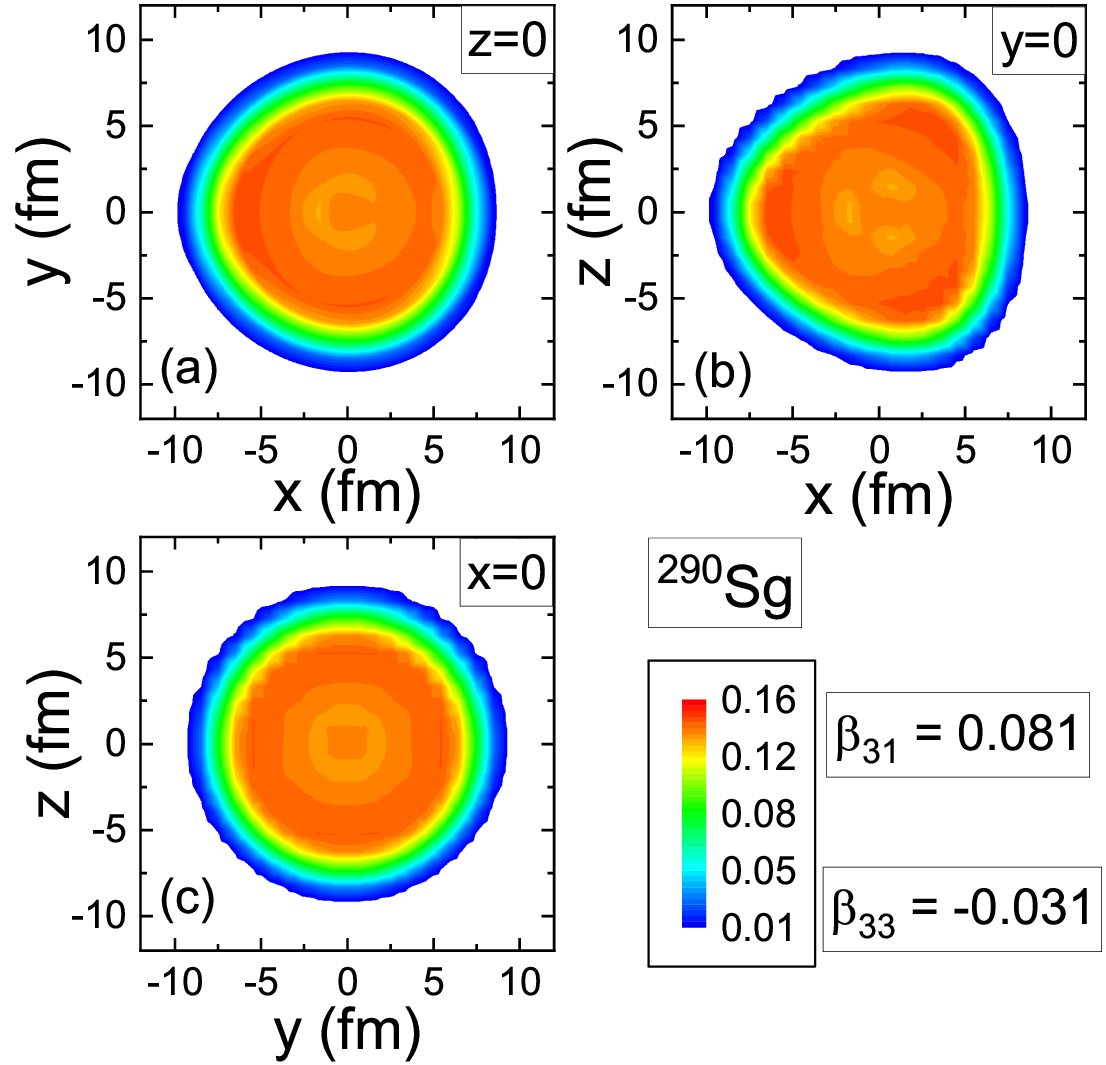}
	\end{center}
	\caption{(Color online) 
		Density profiles of $^{290}$Sg in the $x$-$y$ plane with $z=0$ (a), 
		$x$-$z$ plane with $x=0$ (b), and $y$-$z$ plane with $x=0$ (c).
		The functional DD-PC1 is used in the RHB calculations.
	}
	\label{fig:B31Den}
\end{figure}

In Fig.~\ref{fig:B31Den}, we display the ground state density profiles of $^{290}$Sg
obtained from RHB calculations with the parameter set DD-PC1,
where the nonaxial octupole deformation parameters are predicted to be 
$\beta_{31} = 0.081$ and $\beta_{33} = -0.031$.
Panels (a) and (b) show the density profiles in the $x$-$y$ plane with $z=0$ 
and in the $x$-$z$ plane with $y=0$ respectively. 
Obviously, they are both derivative from a circle, but the degrees of distortion are different. 
For the density profiles in the $y$-$z$ plane with $x=0$ (panel c), the distortion is not visible.

\begin{figure}[h]
	\begin{center}
		\includegraphics[width=0.45\textwidth]{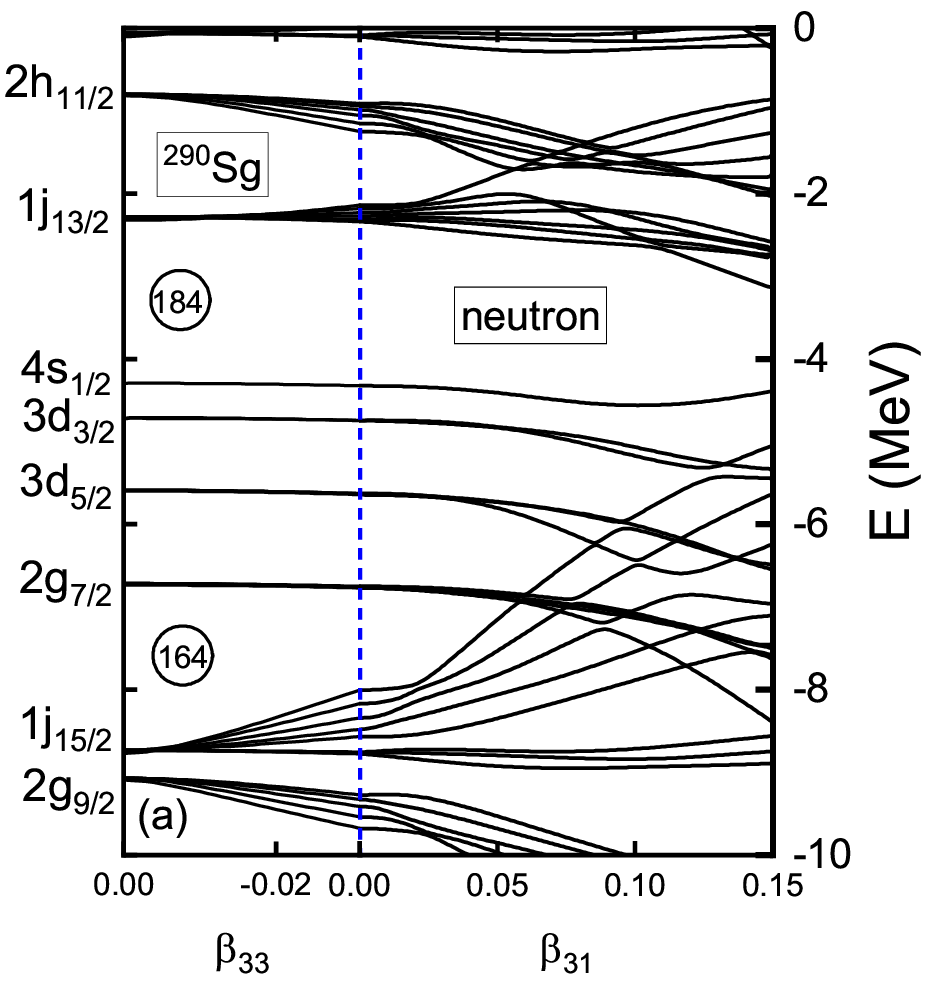} \\
		\includegraphics[width=0.45\textwidth]{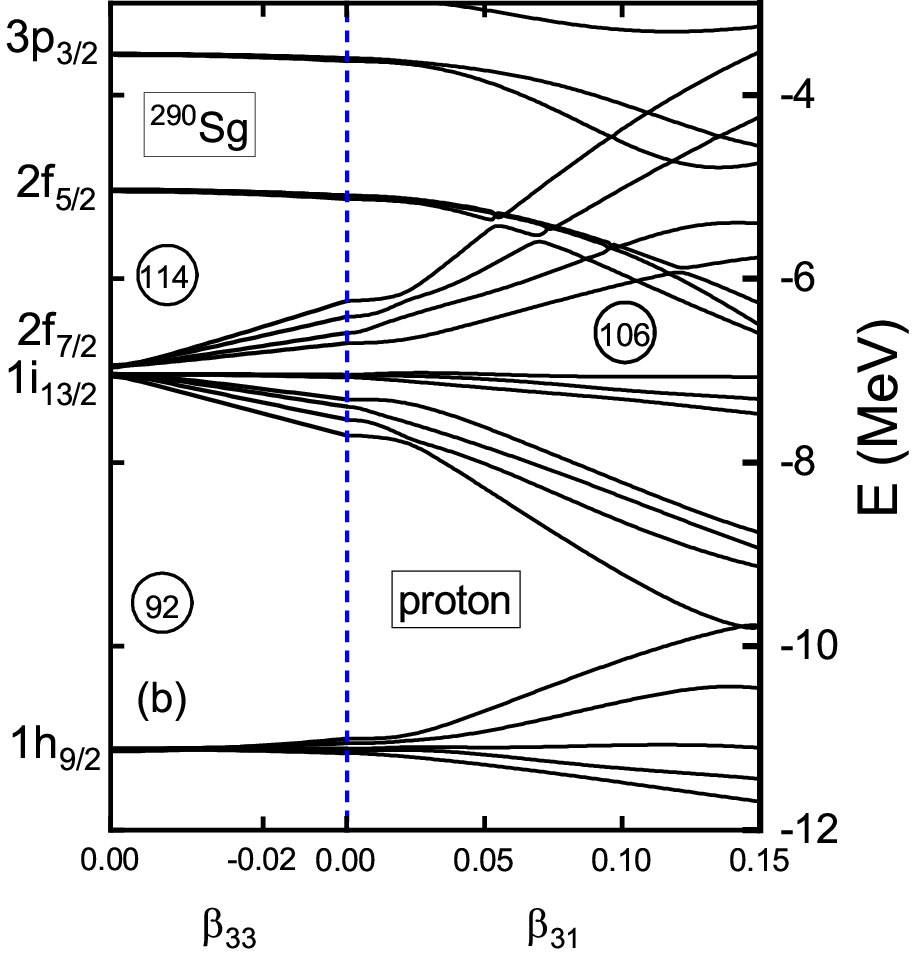}
	\end{center}
	\caption{(Color online)
		The single-particle levels near the Fermi surface for neutrons (upper panel) and protons (lower panel) 
		of $^{290}$Sg as functions of the nonaxial octupole deformation $\beta_{33}$ (left side) and of $\beta_{31}$
		with $\beta_{33}$ fixed at $-0.031$ (right side).
	}
	\label{fig:spl}
\end{figure}

In Fig.~\ref{fig:spl} we show the neutron and proton single-particle levels near the Fermi surface of 
$^{290}$Sg as functions of the nonaxial octupole deformation $\beta_{33}$ on the left side 
and of $\beta_{31}$ with $\beta_{33}$ fixed at $-0.031$ on the right side.
In the upper panel of Fig.~\ref{fig:spl}, a pronounced spherical shell closure at $N=184$ is evident,
well-established by early studies~\cite{Rutz1997_PRC56-238,Gupta1997_MPLA12-1727,
	Patra2000_JPG26-L65,Agbemava2015_PRC92-054310}.
For $\beta_{31} \leq 0.1$, this energy gap remains stable,
gradually decreasing only as $\beta_{31}$ surpasses $0.1$.
In the lower panel, the proton single-particle levels are present.
Notably, the spherical proton orbitals $1i_{13/2}$ and $2f_{7/2}$ are in close proximity,
satisfying the $\Delta j = \Delta l = 3$ condition, 
and their near degeneracy give rise to ocutupole correlations.
With increasing values of $\beta_{33}$ and $\beta_{31}$ from zero, an energy gap emerges at $Z=106$,
while the spherical shell gaps at $Z=92$ and 104 are suppressed.
Due to the substantial energy gap at $Z=106$, a robust nonaxial octupole effect is 
anticipated for $^{290}$Sg and neighboring nuclei.

\begin{table*}
	\centering
	\caption{\label{tab:gro} %
		The quadrupole deformations $\beta_{20}$ and $\beta_{22}$, the octupole deformations 
		$\beta_{3\mu}$, $\mu=0,1,2,3$ and the hexadecapole deformation $\beta_{40}$ together 
		with the binding energies $E_{\rm{cal.}}$ for the ground states of $N=184$ nuclei 
		calculated with parameter sets DD-PC1, DD-ME2, and PC-PK1. $E_{\rm{depth}}$ denotes 
		the energy difference between the ground states and the point constrained $\beta_{3\mu}$
		to zero. All energies are in MeV.
	}
	\footnotesize
	\begin{tabular}{lccccccccccc}
		\hline\hline
		Nucleus	   & Parameters & $\beta_{20}$ & $\beta_{22}$ & $\beta_{30}$ & $\beta_{31}$ & $\beta_{32}$ & $\beta_{33}$ & $\beta_{40}$ & $E_{\rm{cal.}}$ & $E_{\rm{depth}}$ \\ \hline
		$^{284}$Fm & DD-PC1     & 0.005 & 0.002 & 0.000 & 0.077 & 0.000 & 0.042    & 0.003 & $-2027.545$ & 0.779 \\ 
		& DD-ME2     & 0.003 & 0.005 & 0.000 & 0.049 & 0.000 & $-0.016$ & 0.001 & $-2024.425$ & 0.241 \\ 
		& PC-PK1     & 0.001 & 0.001 & 0.000 & 0.029 & 0.000 & $-0.007$ & 0.000 & $-2033.192$ & 0.027 \\ 
		$^{286}$No & DD-PC1     & 0.005 & 0.006 & 0.000 & 0.078 & 0.000 & $-0.016$ & 0.002 & $-2044.183$ & 0.907 \\ 
		& DD-ME2     & 0.003 & 0.002 & 0.000 & 0.065 & 0.000 & $ 0.026$ & 0.002 & $-2041.995$ & 0.514 \\ 
		& PC-PK1     & 0.002 & 0.002 & 0.000 & 0.045 & 0.000 & $-0.012$ & 0.001 & $-2048.964$ & 0.111 \\
		$^{288}$Rf & DD-PC1     & 0.004 & 0.004 & 0.000 & 0.084 & 0.000 & $-0.008$ & 0.003 & $-2059.616$ & 1.062 \\ 
		& DD-ME2     & 0.002 & 0.002 & 0.000 & 0.068 & 0.000 & $ 0.007$ & 0.002 & $-2058.097$ & 0.580 \\ 
		& PC-PK1     & 0.001 & 0.002 & 0.000 & 0.051 & 0.000 & $-0.013$ & 0.001 & $-2063.518$ & 0.161 \\
		$^{290}$Sg & DD-PC1     & 0.002 & 0.003 & 0.000 & 0.081 & 0.000 & $-0.031$ & 0.003 & $-2073.639$ & 1.033 \\ 
		& DD-ME2     & 0.001 & 0.001 & 0.000 & 0.065 & 0.000 & $-0.017$ & 0.002 & $-2072.828$ & 0.529 \\ 
		& PC-PK1     & 0.000 & 0.000 & 0.000 & 0.050 & 0.000 & $-0.013$ & 0.001 & $-2076.818$ & 0.136 \\
		\hline\hline
	\end{tabular}
\end{table*}

To examine the dependence of our results on the functional form and on the effective interaction, 
we also studied $^{284}$Fm, $^{286}$No, $^{288}$Rf, and $^{290}$Sg with parameter sets 
DD-ME2~\cite{Lalazissis2005_PRC71-024312} and PC-PK1~\cite{Zhao2010_PRC82-054319}.
The results are listed in Table~\ref{tab:gro}.
Roughly speaking, the outcomes are similar with different parameter sets, 
i.e., for the ground states, $\beta_{30}$ and $\beta_{32}$ vanish in all cases, 
$\beta_{20}$, $\beta_{22}$, and $\beta_{40}$ are extremely small, thus can be viewed as zero.
For all nuclei investigated here, DD-PC1 predicts the largest energy gain $E_{\rm{depth}} \approx 1$ MeV
with $\beta_{31} \approx 0.08$. 
For DD-ME2, the predicted energy gain due to $\beta_{31}$ distortion is about 0.5 MeV for 
$^{286}$No, $^{288}$Rf, and $^{290}$Sg, while $E_{\rm{depth}} \approx 0.2$ MeV for $^{284}$Fm.
The parameter set PC-PK1 predicts the smallest energy gain, $E_{\rm{depth}} < 0.2$ MeV for all cases; 
especially for $^{284}$Fm, this value is only about 30 keV.
The predicted $\beta_{31}$ values are around 0.05 for $^{286}$No, $^{288}$Rf, and $^{290}$Sg,
while for $^{284}$Fm, the equilibrium $\beta_{31}$ value is $\sim$ 0.03.
The evolution of the nonaxial-octupole $\beta_{31}$ and $\beta_{33}$ effect 
along the $N = 184$ isotonic chain is almost independent of the
form of energy density functional and the parameter set: 
The effect from $\beta_{31}$ and $\beta_{33}$ distortion is the strongest in $^{288}$Rf
and the smallest in $^{284}$Fm.

\section{\label{sec:summary}Summary}
In summary, we studied the ground state shapes of $N=184$ isotones
within the covariant density functional theory.
To investigate the role played by the four octupole deformations $\beta_{3\mu}$,
$\mu=0$, 1, 2,3, in determining the ground state shapes of $N=184$ isotones, 
we solved the Dirac Hartree-Bogoliubov equation in simplex-$y$ 
harmonic oscillator basis.
One-dimensional constraint calculations were performed by constraining each 
$\beta_{3\mu}$, $\mu=0$, 1, 2,3 while other shape degrees of freedom were
determined automatically start from zero initial values.
The most pronounced lowering effect was observed by constraining $\beta_{31}$.
Simultaneously, only the self-consistently determined $\beta_{33}$ value is nonzero at the minimum.
Thus the ground state shapes of $^{284}$Fm, $^{286}$No, $^{288}$Rf, and 
$^{290}$Sg were predicted at $\beta_{31} \approx 0.08$ and 
$\beta_{33} \approx -0.01 \sim -0.03$ with the covariant density functional DD-PC1. 
The lowering effect due to $\beta_{3\mu}$ distortion can be attributed to 
the interaction between the proton orbitals from $1i_{13/2}$ and $2f_{7/2}$, 
as a consequence, large energy gap at $Z=106$ was developed at 
$\beta_{3\mu} \neq 0$ in the single particle levels.
The results from covariant density functional DD-ME2 and PC-PK1 are consistent with
the one from DD-PC1, thus the distortion effect from $\beta_{31}$ and $\beta_{33}$ 
along the N = 184 isotonic chain is not very sensitive to the form of the 
energy density functional and the parameter set we used.
We should note that the present results are limited to the mean-field calculations, 
to accurately clarify the effect of the non-axial octupole deformations on the ground 
and low-lying states of these nuclei, the beyond mean-field calculations, 
for example the full symmetry-restored generator-coordinate method calculations are necessary.
\bigskip
\acknowledgements
This work has been supported by the National Natural Science 
Foundation of China under Grant No. 12005107.


%

\end{document}